# Carbon and vacancy centers in hexagonal boron nitride.

P. Huang[1,2,3], M. Grzeszczyk[1,2], K. Vaklinova[1,2], K. Watanabe[4], T. Taniguchi[5], K. S. Novoselov[1,2], M. Koperski[1,2,*]

[1] Department of Materials Science and Engineering, National University of Singapore, 117575, Singapore
[2] Institute for Functional Intelligent Materials, National University of Singapore, 117544, Singapore
[3] Guangxi Key Laboratory of Information Materials, Guilin University of Electronic Technology, Guilin, 541004, China
[4] Research Center for Functional Materials, National Institute for Materials Science, Tsukuba 305-0044, Japan
[5] International Center for Materials Nanoarchitectonics, National Institute for Materials Science, 305-0044, Japan

E-mail: *msemaci@nus.edu.sg

**Creation of defect with predetermined optical, chemical and other characteristics is a powerful tool to enhance the functionalities of materials. Herewith, we utilize density functional theory to understand the microscopic mechanisms of formation of defects in hexagonal boron nitride based on vacancies and substitutional atoms. Through in-depth analysis of the defect-induced band structure and formation energy in varying growth conditions, we uncovered a dominant role of interdefect electron paring in stabilization of defect complexes. The electron reorganization modifies the exchange component of the electronic interactions which dominates over direct Coulomb repulsion or structural relaxation effects making the combination of acceptor- and donor-type defect centers energetically favorable. Based on an analysis of a large number of defect complexes we develop a simple picture of the inheritance of electronic properties when individual defects are combined together to form more complex centers.**

## Introduction

Controllable formation of defect in crystals is a pathway towards observation of novel phenomena, tailoring material properties and enabling new functionalities**[1]**. Successful formation of specific defects typically requires knowledge about microscopic mechanisms of defect-lattice or interdefect interactions. Defect stability needs to be understood in relation to the electronic properties of the defect-enriched material**[2-10]**. With the recent observation of single photon emitters in hexagonal boron nitride (hBN), the intentional creation of defect centers became a technologically important challenge. hBN is a representative of two-dimensional insulators**[11,12]** hosting a variety of defect centers that activate mid-gap photoluminescence (PL) in ultraviolet, visible and near-infrared spectral regions**[13-19]**. These luminescent centers appear spontaneously in different types of hBN crystals: mechanically exfoliated layers, micro- and nanoscale powders, chemical vapor deposition (CVD) or metal-organic vapor deposition (MOVPE) grown thin films**[20]**. They can also be created and/or enhanced via annealing, irradiation or staining of hBN layers. Some of these defects are currently associated with carbon impurities, which can be introduced in-situ during crystal growth or via post-growth annealing of pristine hBN crystals in carbon-rich atmosphere**[21,22]**.

The optical response of the carbon-doped hBN is likely dominated by the carbon impurities and/or vacancies that can be found in pristine hBN films. The combination of single site vacancies and carbon substitutions creates a vast parameters space to be explored when considering defect-related optical transitions. The complexity of the problem is reflected in the PL response of carbon-doped hBN films**[21-28]**. The defect-induced optical processes are manifested in form of multi-resonance patterns with well-defined energy of transitions (**Fig. 1a**) or appear as more disordered systems characterized by broad-band emission attributed to ensembles of defects (**Fig. 1b**).

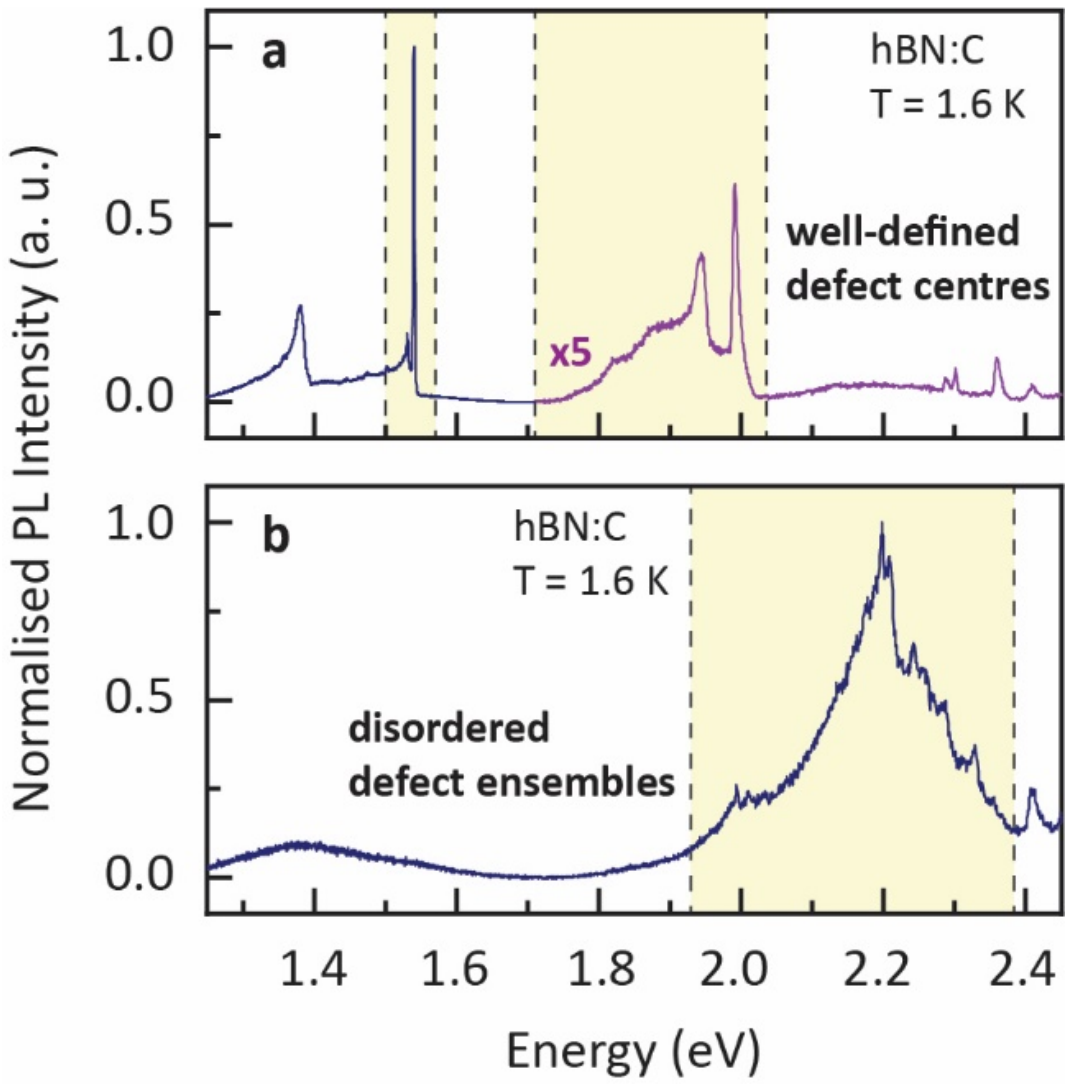

**Figure 1.** Example low temperature (1.6 K) μPL spectra of carbon-doped hBN films characterised by few tens of nanometer thickness. **(a)** A reproducible pattern of optical resonances with well-defined energy of transitions is the most common optical response. **(b)** Some crystals are characterised by emission spectra that are dominated by broader PL bands composed of a series of narrow lines indicative of more disordered defect ensemble.

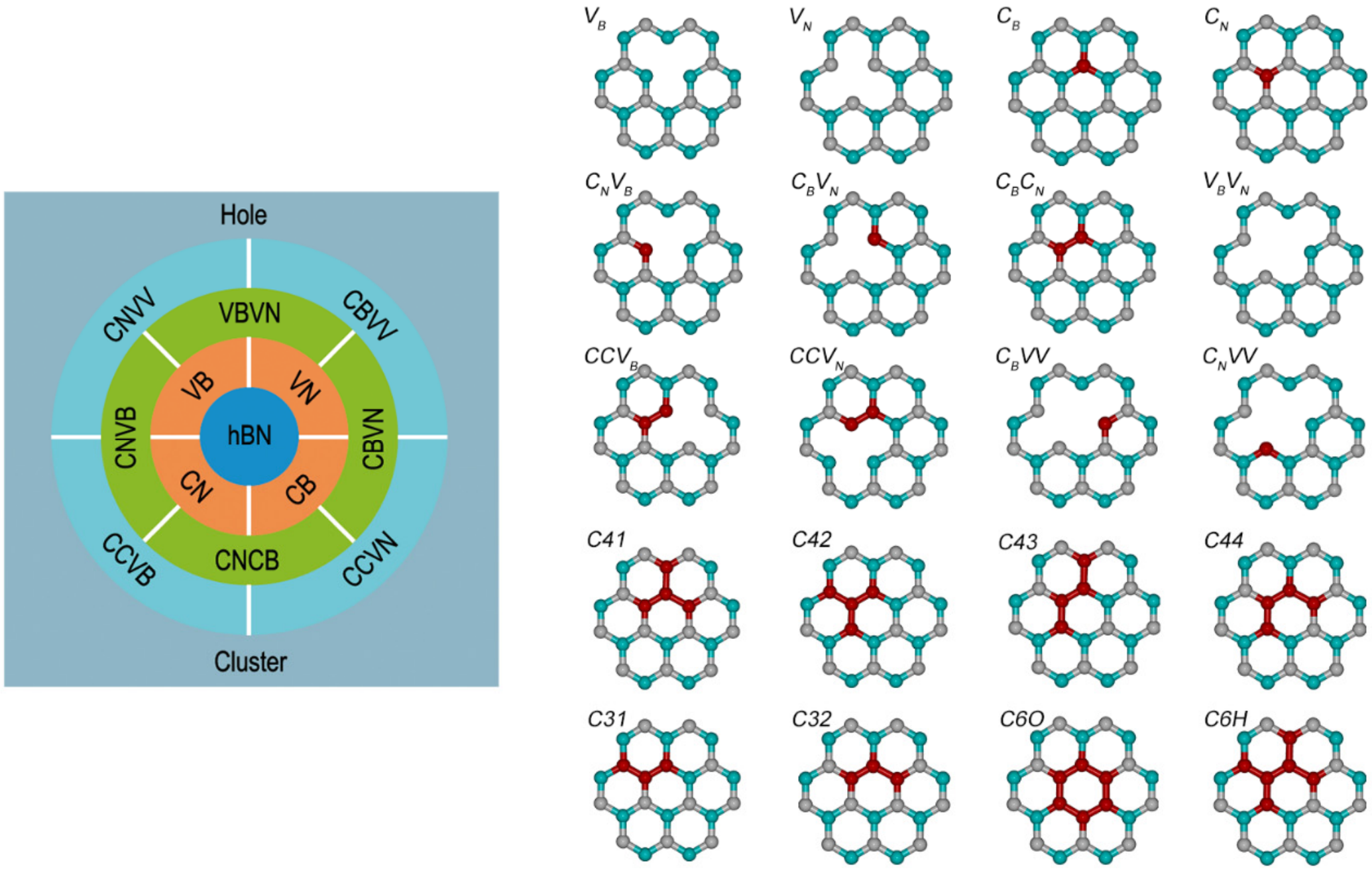

**Figure 2.** The atomic structure of the vacancy and carbon defect centres considered in current work. Such defect library is characterised by hierarchical structure, when more complex lattice imperfections can be formed from simpler constituent defects. Merging of the defects may be investigated from the point of view of the lattice stability or inheritance of the electronic properties.

In order to understand the probability of defect formation in various experimental conditions, we investigated systematically the electronic properties and stability of carbon and vacancy centers in hexagonal boron nitride. We can classify these centers by their complexity ranging from single site substitutions and vacancies to clusters of defects as illustrated schematically in **Fig. 2**. Through a comparative analysis of the defect properties, we have found that it is energetically favorable to couple two elementary defects together through a mechanism of interdefect electron pairing. The process of combining defects resembles formation of interatomic chemical bonds, determined by the competition between the Coulomb and exchange interactions arising from overlapping atomic orbitals**[29,30]**. We demonstrate that within hBN lattice combination of carbon and/or vacancy centers introduces elegant symmetries within the electronic band structure, leading to robust stabilization of the defect complexes. The mechanism is driven by the modification of the exchange terms, which dominate over the increased Coulomb repulsion.

## Single and double site defect centers

Pristine hBN lattice is composed of van der Waals bonded monolayers in AA' stacking configuration. A monolayer hBN is characterized by hexagonal structure with two triangular sub-lattices occupied by boron and nitrogen. The existence of two different atoms in the elementary cell breaks the symmetry of the hexagonal lattice, which leads to an opening of the band gap of about 6.0 eV in the bulk limit. The in-plane interatomic $\sigma$-type bonding between boron and nitrogen arises from $sp^2$ hybridization. It is predominantly responsible for the structural stability of the crystal. The fundamental valence and conduction band are constructed from $\pi$ bonded and $\pi^*$ anti-bonded orbitals, respectively.

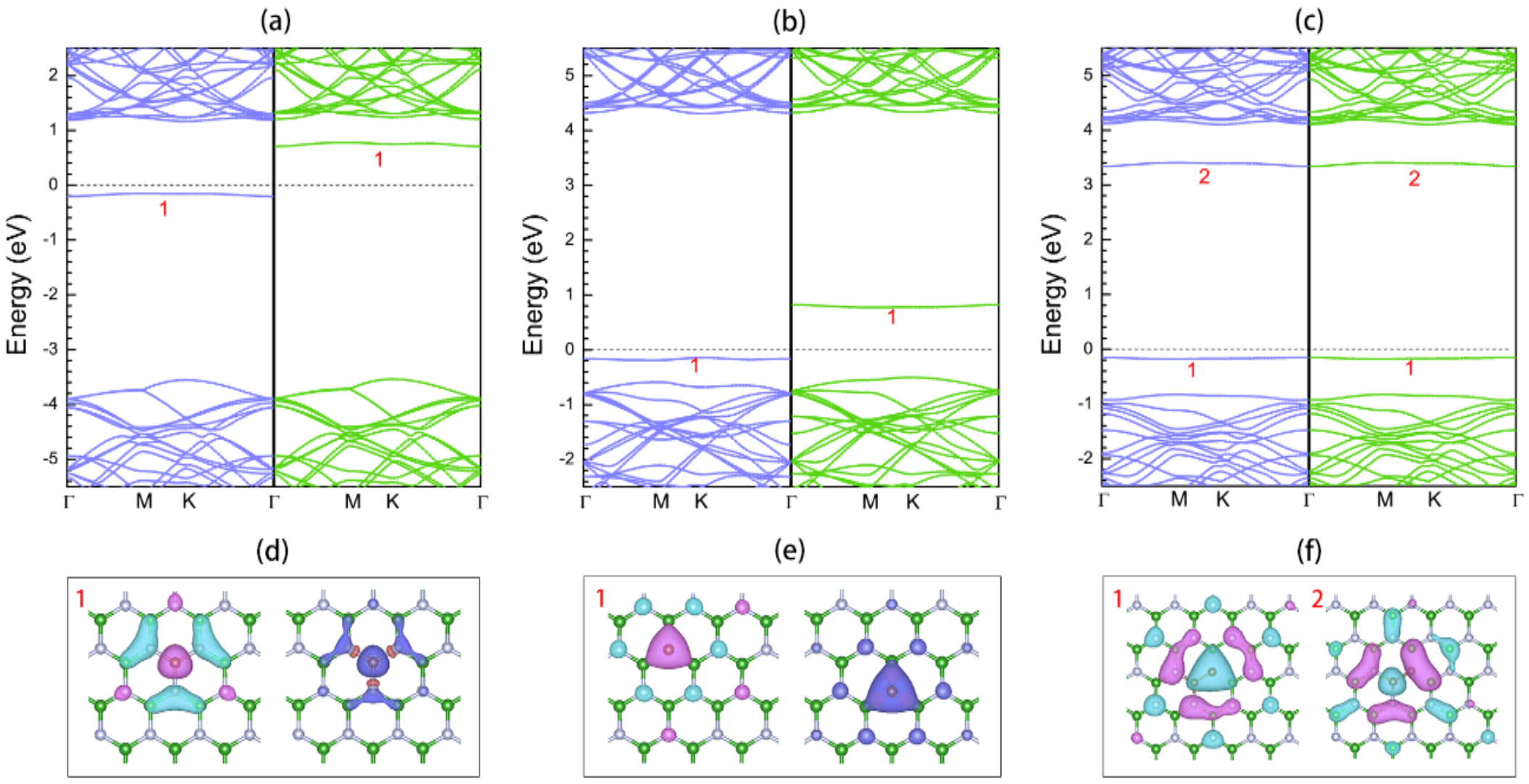

**Figure 3.** Spin-projected band structure of hBN in the presence of carbon substitution centers: **(a)** $C_B$, **(b)** $C_N$ and **(c)** $C_BC_N$. **(d-f)** The top view of $\Gamma$-point wave functions (real part) illustrate the defect states marked by numbers (1 or 2) in the corresponding band diagrams. The colour coding (pink, light blue) represents opposite signs of the wave function. **(d,e)** The total spin density is presented with colour coding (purple, brown) depicting spin up and down states.

Carbon doping of hBN leads to the creation of elemental impurities such as carbon substitution for boron ($C_B$) or carbon substitution for nitrogen ($C_N$). Both defect centers give rise to a spin-split midgap level. The defect-induced level is located below the conduction band for $C_B$ creating a donor state or above the valence band for $C_N$ creating an acceptor state as illustrated in **Fig. 3(a,b)**. Combining these two defects together to form a $C_NC_B$ dimer impacts qualitatively the electronic structure of the defect. $C_NC_B$ can be considered as a simple molecule embedded within hBN lattice with energy levels defined by $\pi$-bonding between $p_z$ orbitals of the C atoms at N and B sites. As a result of the C-C bonding the spin-splitting is lifted via reorganization of the electrons from occupying $C_N$ and $C_B$ states to occupying a molecular-like $C_BC_N$ level. Consequently, the electronic structure of $C_BC_N$ defect resembles HOMO[1] and LUMO[2] levels residing above the valence band edge and below conduction band edge, respectively. The HOMO level is fully occupied and the LUMO levels is empty, leading to removal of the spin-spilling observable for single site C-defects.

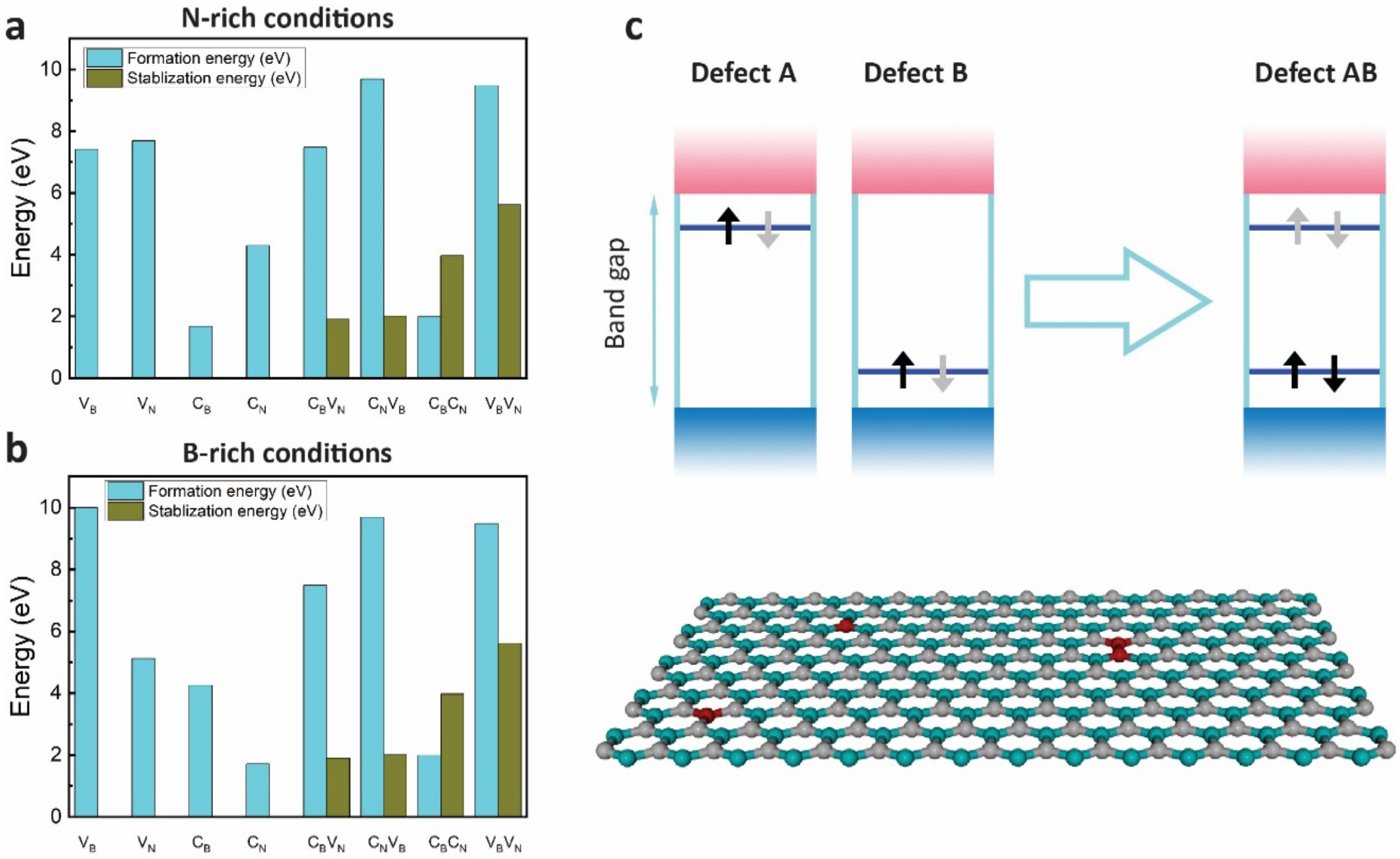

**Figure 4.** The histograms represent formation energy for single site and double site defect centres in **(a)** nitrogen-rich and **(b)** boron-rich condition together with the absolute value of the stabilization energy for double-site defects. **(c)** An schematic diagram illustrates a process of combining two defects that give rise to half-occupied donor and acceptor states. The combination results in a defect that displays molecular like occupied and empty levels. The mechanism provides a significant contribution to the large values of the stabilisation energy observed for more complex vacancy and carbon defect centers in hBN.

Interestingly, the electronic configuration of B, C and N in combination with the electronic band structure of hBN leads to unique stabilization mechanisms for carbon and vacancy centres. We investigate the formation energy of single site defects and their double site combinations by density functional theory (DFT) for nitrogen- and boron-rich conditions (see **Methods** section for the detailed computational procedure). We define the stabilization energy of multisite defects as the differential formation energy between the defect complex and the sum of the constituent single-site defects:

$$E_{stab}(D) = E_f(D) - \sum_i E_f(D_i)$$

where $E_{stab}(D)$ is the stabilization energy for defect *D*, $E_f(D)$ is the formation energy for defect *D* and $E_f(D_i)$ is the formation energy of single-site constituent defects $D_i$ which combine to form defect

[1] Highest occupied molecular orbital

[2] Lowest unoccupied molecular orbital

*D*. We found that the stabilization energy is negative for all investigated vacancy- and carbon-based defect complexes as illustrated in **Fig. 4(a,b)**. This observation indicates that it is energetically favorable to combine simple defects within hBN lattice to form more complex lattice imperfections. In carbon-doped hBN the dominant mechanism of the reduction of formation energy originates from the reorganization of the electronic structure related to the bond formation. When combining $C_N$ and $C_B$ defect to form the $C_BC_N$ dimer, the electron occupying the $C_B$ donor state relaxes onto the hybridized $C_BC_V$ state. The wave functions of the corresponding states are presented in **Fig. 3(d-f)**. That process dominates over the Coulomb repulsion originating from placing two electrons on the same molecular level and over the energy modification due to defect-induced structural relaxation. Such robust stabilization mechanism via electronic reconfiguration occurs due to the large energy separation of the donor and acceptor states enabled by the hBN band gap. A schematic illustration of the generalized mechanism is presented in **Fig. 4(c)** and the exact expressions for the competing exchange and Coulomb terms are discussed in the Supplementary Information Appendix.

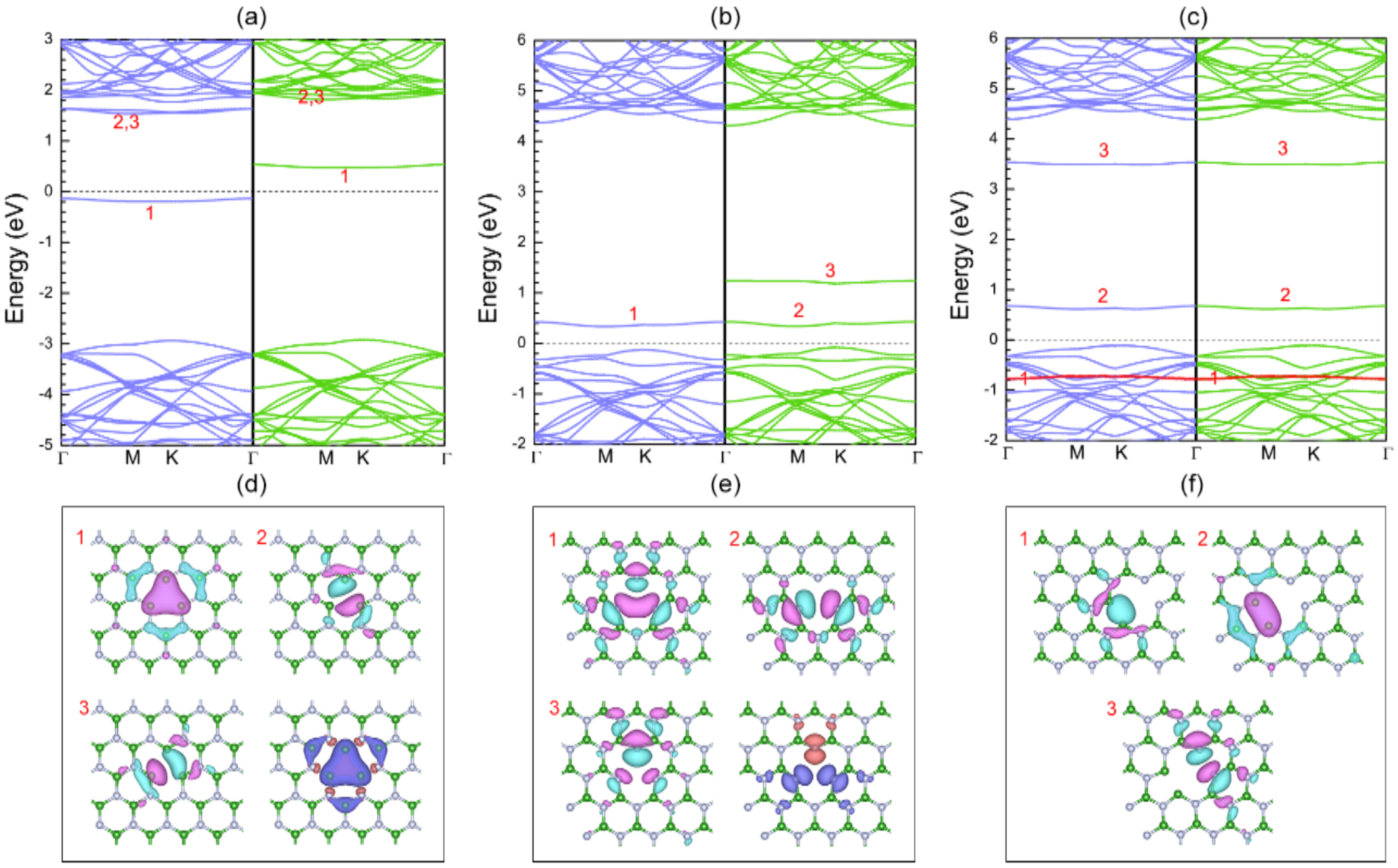

**Figure 5.** Spin-projected band structure of hBN in the presence of vacancy centers: **(a)** $V_N$, **(b)** $V_B$ and **(c)** $V_BV_N$. The red line in **(c)** highlights the defect state of $V_BV_N$ located within the valence band. **(d-f)** The top view of $\Gamma$-point wave functions (real part) illustrate the defect states marked by numbers (1,2 or 3) in the corresponding band diagrams. The colour coding (pink, light blue) represents opposite signs of the wave function. **(d,e)** The total spin density is presented with colour coding (purple, brown) depicting spin up and down states.

A qualitatively similar mechanism is also applicable to the combination of two vacancies in hBN. The formation energy for a nitrogen ($V_N$) or boron ($V_B$) vacancies is larger than for carbon substitutions at corresponding lattice sites ($C_N$ or $C_B$) as they require breaking of three strong $sp^2$ bonds. The $V_N$ defect gives rise to mid-gap states akin to $C_B$. Half-filled level emerges below the conductions band with an addition of two more empty defect levels located within the conduction band as seen in **Fig 5(a)**. The $V_B$ defect creates three half-filled levels. The occupied states are buried within the valence band, while the empty states emerge above the valence band as illustrated in **Fig. 5(b)**. These states originate from the dangling bonds of B atom for $V_N$ defect and from the dangling bonds of the N atom for the $V_B$ defect. The orbital wave functions for these states are visualized in **Fig. 5(d-f)**. The stabilization energy for $V_BV_N$ defect is larger than for $C_BC_N$ defect. The $V_BV_N$ defect creates three molecular-like levels, however the occupied state is located within the valence band, so that the relaxation of the electron from $V_N$ states leads to the increase of the stabilization energy. Additionally, the combination of two vacancies together reduces the number of the broken bonds from six to four. Overall, in the combined $V_BV_N$ state, the three defect-induced levels of the following origin: 1) the empty state below conduction band arises due to out-of-plane $\pi$-bonded $p_z$ orbitals of B-B pair, 2) the empty states above the valence band arises due to in-plane $\sigma^*$ bond of the N-N pair, 3) the occupied state located within the valence band arises due to in-plane bond of the B-B pair.

## Triple site defects and carbon clusters

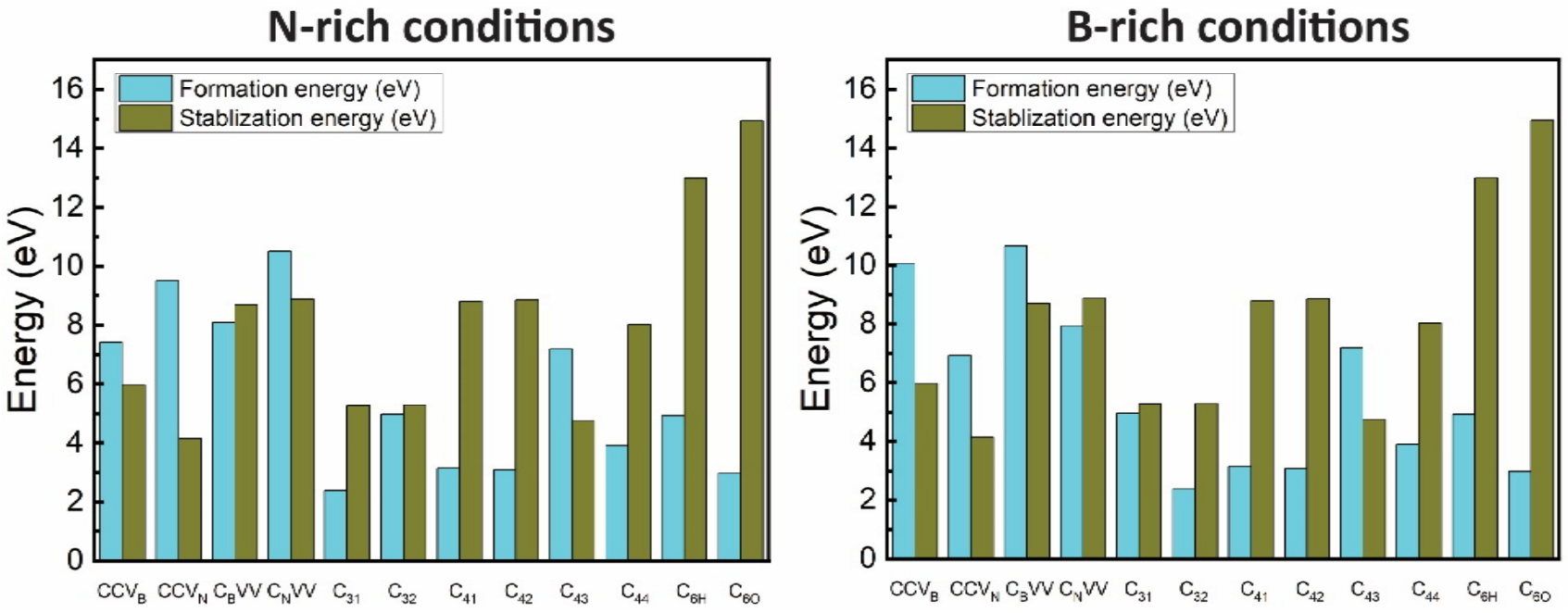

**Figure 6.** The histograms present the formation and stabilization energy for triple site carbon and vacancy centres as well as carbon clusters in hBN in nitrogen-rich and boron-rich conditions.

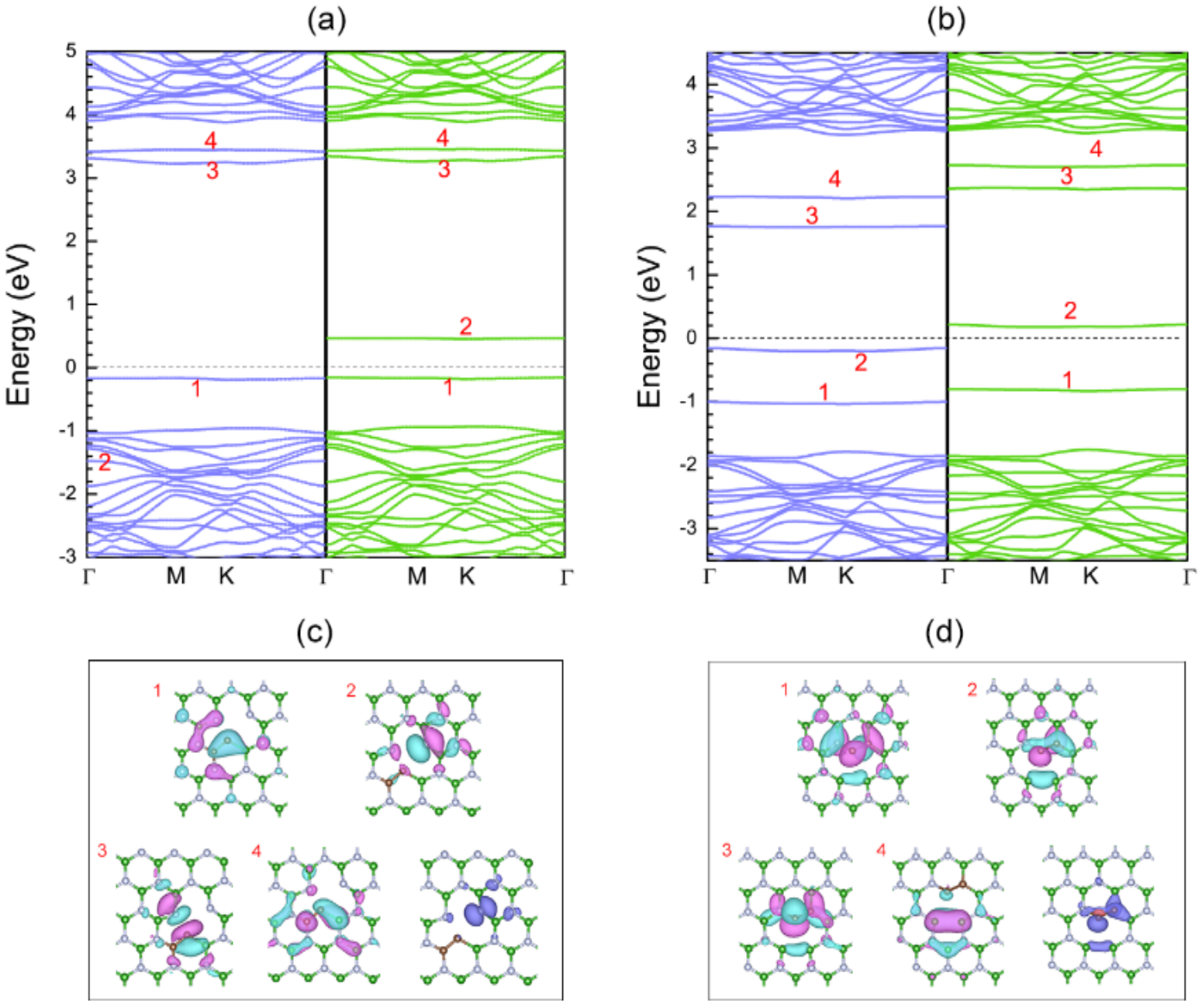

**Figure 7.** Spin-projected band structure of hBN in the presence of triple site centers: **(a)** $C_BC_NV_B$ and **(b)** $C_BC_NV_N$. **(c-d)** The top view of $\Gamma$-point wave functions (real part) illustrate the defect states marked by numbers (1,2, 3 or 4) in the corresponding band diagrams. The colour coding (pink, light blue) represents opposite signs of the wave function. **(c,d)** The total spin density is presented with colour coding (purple, brown) depicting spin up and down states.

The trend of observing negative stabilization energy persists for triple site defects as demonstrated in **Fig. 6**. The triple site defect in carbon-doped hBN may be formed by combining the carbon dimer with a vacancy, vacancy dimer with a single carbon substitution or by creating vacancy or carbon trimers. Here, we will focus on inheritance effects when a carbon dimer is merged with a vacancy. As demonstrated in **Fig. 7**, $C_BC_NV_B$ defect gives rise to four mid-gap states of different origin. The fully occupied state above the valence band emerges from $\pi$-bonding of the C-C dimer and can be considered as an inherited HOMO state. The half-filled state above the valence band originates from the dangling bond of the N atom created by the vacancy. Additionally, two empty states below the conduction band arise due to in-plane $\sigma$* C-B antibonding enabled by the presence of the vacancy (lower energy state) and $\pi$* out of plane C-C bonding (higher energy state). Generally speaking, $C_BC_NV_B$ creates states that are inherited from the C-C dimer and from the vacancy as well as qualitatively new states due to the presence of a vacancy next to a carbon substitution.

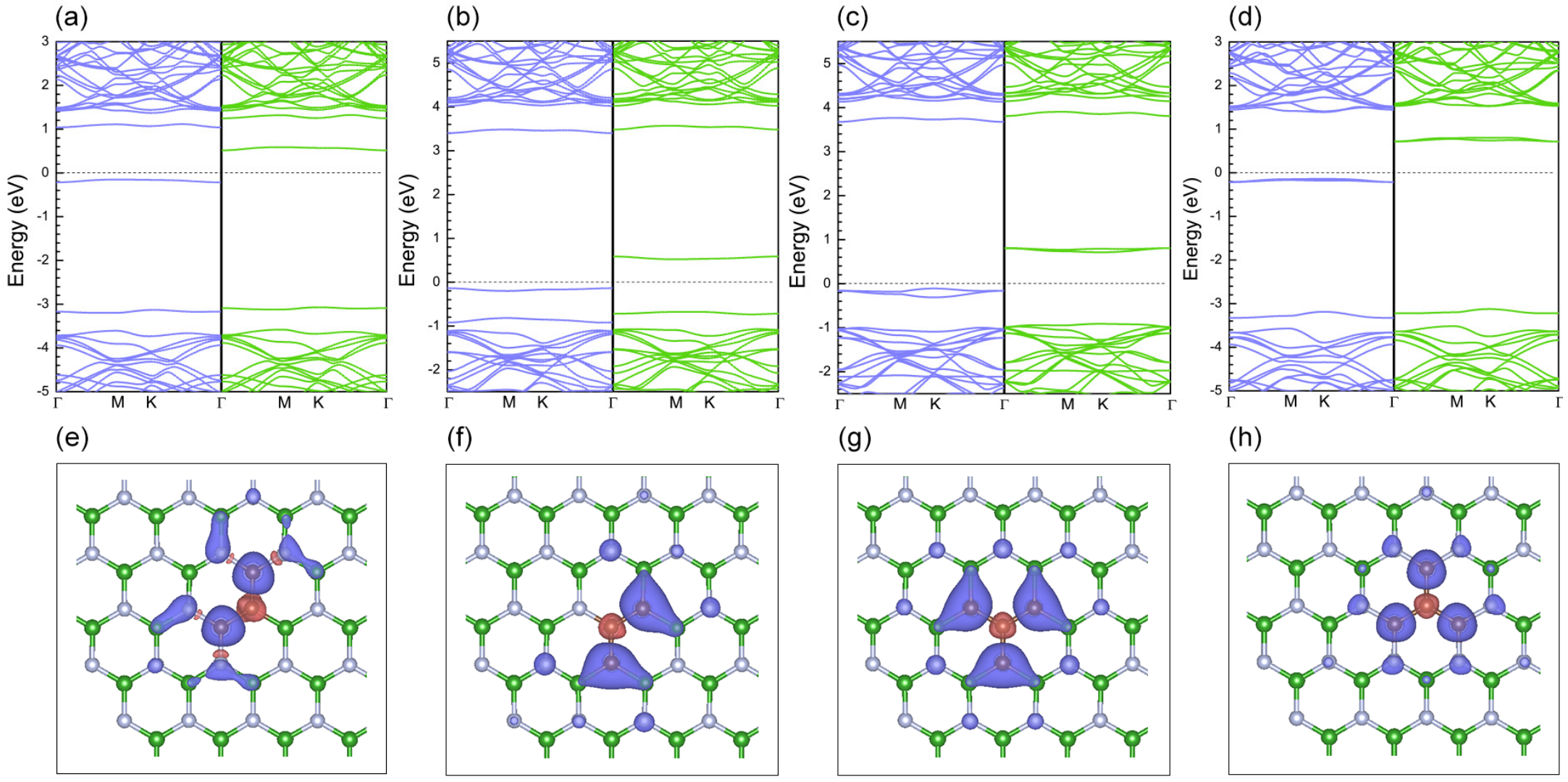

**Figure 8.** Spin-projected band structure of hBN in the presence of carbon clusters: **(a)** $C_{31}$, **(b)** $C_{32}$, **(c)** $C_{41}$ and **(d)** $C_{42}$. The atomic structure of these clusters is presented in **Fig. 2**. **(e-h)** The total spin density is presented with colour coding (purple, brown) depicting spin up and down states.

$C_BC_NV_N$ defect also creates four mid-gap states that can be associated predominantly with carbon dimer and nitrogen vacancy. However, in this case the structural relaxation leads to a displacement of the $C_B$ atom out of the plane of the monolayer hBN. The lattice reconfiguration modifies the orbital composition of the states that can be visualized by the wave functions presented in **Fig. 7(c,d)**. The alteration of the orbital states can be seen in the electronic band structure as the emergent spin-splitting of all four levels as well renormalization of the energy of the levels.

The donor-acceptor pairing mechanism is also relevant for the formation of carbon clusters, which we investigate based on triple, quadruple and sextuple carbon substitutions arranged in various configurations within hBN lattice. The major finding is that every $C_NC_B$ pair within the cluster significantly decreases the formation energy. As a consequence, the quadruple carbon clusters with equal number of $C_B$ and $C_N$ ($C_{41}$ and $C_{42}$) are characterized by lower formation energy than quadruple carbon clusters with imbalanced number $C_B$ and $C_N$ ($C_{43}$ and $C_{44}$) and triple carbon substitutions ($C_{31}$). Among investigated clusters, $C_{6O}$ (an individual carbon hexagonal ring) displays largest stabilization energy, both in absolute terms and by value normalized per carbon atom. The large stabilization energy of $C_{6O}$ originates from the electron pairing in combination with the delocalized character of the $\pi$ orbitals within the carbon ring akin to that of a benzene molecule, leading to significant decrease of the kinetic energy of electrons.

From the point of view of the electronic structure, carbon clusters inherit the states mostly from C-C dimers, $C_B$ and $C_N$ defects. As in previously discussed cases, carbon dimers introduce empty of fully occupied molecular-like states while excess substitutions give rise to half-filled states above and/or below conduction band. As a general rule, the total spin associated with the cluster is given by the difference between $C_B$ and $C_N$ lattice sites:

$$S = \frac{|N_{C_N} - N_{C_B}|}{2}$$

Further examples of the electronic structure for carbon clusters can be found in **Supplementary Information** Appendix.

## Conclusions

We have investigated the stability and formation mechanisms of defect centers in hBN based on carbon substitution and vacancies within DFT framework. The band structure of hBN in combination with the electronic configuration of valence electrons of carbon, nitrogen and boron atoms gives rise to a unique stabilization mechanism through merging of unpaired electrons created by donor and acceptor centers. The proposed mechanism enables a simple understanding of the defect-modified electronic band structure evolving from simple single site defect towards increasing more complex dimers, trimers and clusters. Our findings are relevant for development of novel growth techniques of defect-enriched hBN crystals, post-growth defect creation methods and engineering of atomically thin optoelectronic devices operating via defect-induced optical processes.

## Acknowledgements

This project was supported by the Ministry of Education (Singapore) through the Research Centre of Excellence program (grant EDUN C-33-18-279-V12, I-FIM), AcRF Tier 3 (MOE2018-T3-1-005). This material is based upon work supported by the Air Force Office of Scientific Research and the Office of Naval Research Global under award number FA8655-21-1-7026. K.W. and T.T. acknowledge support from the Elemental Strategy Initiative conducted by the MEXT, Japan (Grant Number JPMXP0112101001) and JSPS KAKENHI (Grant Numbers JP19H05790 and JP20H00354). P.H acknowledges the National Natural Science Foundation of China (51801041) and scholarship from the Guangxi Education Department (China). The computational work for this article was performed on resources at the National Supercomputing Centre, Singapore.

### Methods

**Crystal growth and sample preparation.** Pristine, ultrahigh purity hBN crystals were grown by high pressure temperature gradient method. The carbon doping was introduced by annealing of pristine crystals in a graphite furnace at 2000 °C for approximately an hour. The hBN:C films were isolated through mechanical exfoliation of bulk crystals onto $Si/SiO_2$ substrates. In the transfer process, the substrates were heated to 50 °C to increase the yield of laterally large hBN:C flakes, which exhibit homogeneous and reproducible optical emission. The thickness of the films (typically ranging between few layers to few tens of nanometers) was determined by optical force microscopy.

**Optical spectroscopy.** The PL spectra were measured in back-scattering microscopic configuration under 488 nm laser illumination. The sample was cooled down to 1.6 K via dry system with helium exchange gas acting as the cooling agent for the sample. The $Si/SiO_2$ substrates with hBN:C were mounted onto piezo-stages that allow the movement of the sample and positioning the flakes of interest under the microscopy. The PL signal is collected through the same microscopy and collected through a multimode fiber. The emitted light is dispersed by 0.75 m spectrometer and detected by a charge couple device camera.

**DFT calculations.** Our calculations are based on density functional theory (DFT) using the PBE functional as implemented in the Vienna Ab Initio Simulation Package (VASP)**[31-33]**. The interaction between the valence electrons and ionic cores is described within the projector augmented (PAW) approach with a plane-wave energy cutoff of 500 eV[34]. Spin polarization was included for all the calculations. The monolayer of hBN and defects calculations were performed using a 50-atom 5x5 supercell, and the Brillouin zone was sampled using a (12x12x1) Monkhorst-Pack grid. A 15 Å vacuum space was used to avoid interaction between neighboring layers. In the structural energy minimization, the atomic coordinates are allowed to relax until the forces on all the atoms are less than 0.01 eV/Å. The energy tolerance is $10^{-6}$ eV.

The formation energy, i.e., the energy required to create a defect center is given by

$$E_f(D) = E(D) - E(hBN) + \sum_i n_i(\mu_i + E_i)$$

Where $E(D)$ is the total energy of the defect center, $E(hBN)$ is the energy of the host hBN, $n_i$ is the number of an element (B, N, or C) transferred from the supercell to the reservoirs, and $\mu_i$ is the chemical potential defined with respect to the reservoir, i.e., the energy required to move the element from the lattice to the reservoirs, $E_i$ is the energy of the element in solid/gas phase. Herein, we chose solid boron, $N_2$ molecule, and graphene to calculate the energy for B, N, and C respectively. The value of $\mu_B$ and $\mu_N$ are defined by the growth conditions. There is however a constrain on the sum chemical potential: $\mu_B + \mu_N = \mu_{BN}$ where $\mu_{BN}$ is the chemical potential of a BN pair established with respect to the reservoirs. This condition entails an upper bound for the chemical potential of N at the N-rich (B-poor) condition, $\mu_N = 0$ , where the chemical potential of N is in equilibrium with the reservoir. Similarly, the B-rich (N-poor) condition is defined as $\mu_B = 0$. The calculated formation energy is dependent on the chemical potential of the element especially for non-stoichiometric structures.

*Supplementary Information* for **Carbon and vacancy centers in hexagonal boron nitride**.

P. Huang[1,2,3], M. Grzeszczyk[1,2], K. Vaklinova[1,2], K. Watanabe[4], T. Taniguchi[5], K. S. Novoselov[1,2], M. Koperski[1,2,*]

[1] Department of Materials Science and Engineering, National University of Singapore, 117575, Singapore

[2] Institute for Functional Intelligent Materials, National University of Singapore, 117544, Singapore

[3] Guangxi Key Laboratory of Information Materials, Guilin University of Electronic Technology, Guilin, 541004, China

[4] Research Center for Functional Materials, National Institute for Materials Science, Tsukuba 305-0044, Japan

[5] International Center for Materials Nanoarchitectonics, National Institute for Materials Science, 305-0044, Japan

E-mail: *msemaci@nus.edu.sg

**Quantum mechanical origin of the stabilization process for defect complexes in hBN.**

The origin of the stabilization mechanism for the formation of defect complexes in hBN stems from the competition between the exchange interaction and Coulomb repulsion between electrons occupying the bonded interdefect states. When two elementary defects are spatially separated from each other, the correlation between the electrons residing at defect levels is negligible. In such a case, the many-body wave function $\Phi_0$ can be expressed as two independent multiplicative components. As an example, we can consider two isolated carbon substations at boron and nitrogen sites to write the wave function as:

$$\Phi_0(r_1,r_2) = \frac{1}{\sqrt{2}}\phi_N(r_1)\phi_B(r_2)$$

where $\phi_B$ and $\phi_N$ are the single defect orbitals for $C_B$ and $C_N$, respectively. These wave functions fulfil the Schrödinger equations:

$$(T_1 - V_N^1)\phi_N = E_1\phi_N$$

$$(T_2 - V_B^2)\phi_B = E_2\phi_B$$

where $T$ is the kinetic energy of the electron at the defect level and $V_N$ and $V_B$ represent the Coulomb potential energy at the N and B sites contributed by the atomic nuclei and band electrons.

When the two substitutions are combined to occupy the neighboring lattice sites, the Hamiltonian of the system needs to be modified to account for inter-defect interactions:

$$H = (T_1 + T_2) - V_B^1 - V_N^1 - V_B^2 - V_N^2 + \frac{e^2}{4\pi\varepsilon\varepsilon_0 \cdot r_{12}} + \frac{e^2}{4\pi\varepsilon\varepsilon_0 \cdot R_{NB}}$$

Apart from additive single-defect terms, the Hamiltonian includes the Coulomb potential originating from the neighboring defect sites (represented by $V_N^2$ and $V_B^1$) as well as electron-electron and core-core electrostatic interactions in the screening environment of band electrons (described by the dielectric constant $\varepsilon$).

The many-body state arising from such Hamiltonian takes form:

$$\Phi(r_1,r_2) = \frac{1}{\sqrt{2(1 \pm S^2)}}[\phi_N(r_1)\phi_B(r_2) \pm \phi_N(r_2)\phi_B(r_1)]$$

where $S = \langle \phi_B | \phi_N \rangle$ is the overlap integral. The positive and negative signs correspond to the bonding and anti-bonding states, respectively. The many-body state is enriched with an additional term arising due to the exchange of electrons between the N and B substitution sites. The energy of the bonded ($E_+$) and anti-bonded ($E_-$) state can be therefore expressed as:

$$E_{\pm} = \langle \Phi(r_1, r_2) | H | \Phi(r_1, r_2) \rangle = E_0 + \frac{K \pm J}{\sqrt{1 \pm S^2}}$$

where $E_0$ is the sum of single defect energies, K is the Coulomb energy given by:

$$K = \left\langle \phi_N(r_1)\phi_B(r_2) \middle| -V_B^1 - V_N^2 + \frac{e^2}{4\pi\varepsilon\varepsilon_0 \cdot r_{12}} + \frac{e^2}{4\pi\varepsilon\varepsilon_0 \cdot R_{NB}} \middle| \phi_N(r_1)\phi_B(r_2) \right\rangle$$

The Coulomb energy takes into account the electron-core attraction, electron-electron-repulsion and core-core repulsion. Such electrostatic interaction is usually dominated by the repulsive forces, however the exact contribution of each term depends on the type of the defect centres. For $C_NC_B$ defect complex, the on-site electron-electron repulsion provides the strongest contribution and constitutes the dominating term for Coulomb energy.

The J parameter is the exchange integral:

$$J = \left\langle \phi_N(r_2)\phi_B(r_1) \middle| -V_B^1 - V_N^2 + \frac{e^2}{4\pi\varepsilon\varepsilon_0 \cdot r_{12}} + \frac{e^2}{4\pi\varepsilon\varepsilon_0 \cdot R_{NB}} \middle| \phi_N(r_2)\phi_B(r_1) \right\rangle$$

The increased Coulomb repulsion in the bonded state for $C_NC_B$ dimer is compensated by the modification of exchange integral leading to the large value of the stabilization energy as described in the main text.

**Summary of basic properties of carbon and vacancy centers in hBN.**

| **Defects** | **Formation energy (eV)** | | **Stabilization energy (eV)** | **Magnetic moment (μB)** | **Symmetry** |
|---|---|---|---|---|---|
| | N rich condition | B rich condition | | | |
| **$V_B$** | 7.41（7.40） | 9.99 | | 1.0 | $C_{2v}$ |
| **$V_N$** | 7.70（7.70） | 5.12 | | 1.0 | $D_{3h}$ |
| **$C_B$** | 1.68（1.68） | 4.26 | | 1.0 | $D_{3h}$ |
| **$C_N$** | 4.29（4.30） | 1.71 | | 1.0 | $D_{3h}$ |
| | | | | | |
| **$C_BV_N$** | 7.48 | 7.8 | -1.90 | 0.0 | $C_{1h}$ |
| **$C_NV_B$** | 9.69 | 9.69 | -2.00 | 2.0 | $C_{2v}$ |
| **$C_BC_N$** | 2.00 | 2.00 | -3.96 | 0.0 | $C_{2v}$ |
| **$V_BV_N$** | 9.48 | 9.48 | -5.63 | 2.0 | $C_{2v}$ |
| | | | | | |
| **$CCV_B$** | 7.48 | 10.06 | -5.97 | 1.0 | $C_1$ |
| **$CCV_N$** | 9.51 | 6.93 | -4.15 | 1.0 | $C_1$ |
| **$C_BVV$** | 8.08 | 10.66 | -8.7 | 1.0 | $C_1$ |
| **$C_NVV$** | 10.51 | 7.93 | -8.88 | 1.0 | $C_1$ |
| | | | | | |
| **$C_{31}$** | 2.38 | 4.96 | -5.27 | 1.0 | $C_{2v}$ |
| **$C_{32}$** | 4.97 | 2.38 | -5.29 | 1.0 | $C_{2v}$ |
| **$C_{41}$** | 3.15 | 3.15 | -8.79 | 0.0 | $C_1$ |
| **$C_{42}$** | 3.08 | 3.08 | -8.86 | 0.0 | $C_1$ |
| **$C_{43}$** | 7.19 | 7.19 | -4.75 | 2.0 | $D_{3h}$ |
| **$C_{44}$** | 3.91 | 3.91 | -8.03 | 2.0 | $D_{3h}$ |
| **$C_{6(H)}$** | 4.93 | 4.93 | -12.98 | 0.0 | $C_{2v}$ |
| **$C_{6(O)}$** | 2.98 | 2.98 | -14.93 | 0.0 | $D_{3h}$ |

**Table S1.** Calculated formation and stabilization energy, magnetic moment together with corresponding symmetry groups for carbon and vacancy centers in hBN.

**Further examples of the electronic structure of carbon clusters in hBN.**

The gap states of carbon clusters in hBN are essentially the assembly and recombination of that of $C_B$ and $C_N$. Each $C_B$ contributes a gap state composed of $p_z$ orbital below the conduction band while each $C_N$ gives rise to a state composed of $p_z$ orbital above the valence band. For carbon clusters with equivalent B and N site substitutions, including $C_{43}$, $C_{44}$, $C_{6O}$ and $C_{6H}$ structures (see **Fig. 2** in the main text), the gap states are derived from $C_BC_N$ dimers as demonstrated in **Fig. S1**. In such atomic configurations, there are no unpaired electrons resulting in vanishing magnetic moment.

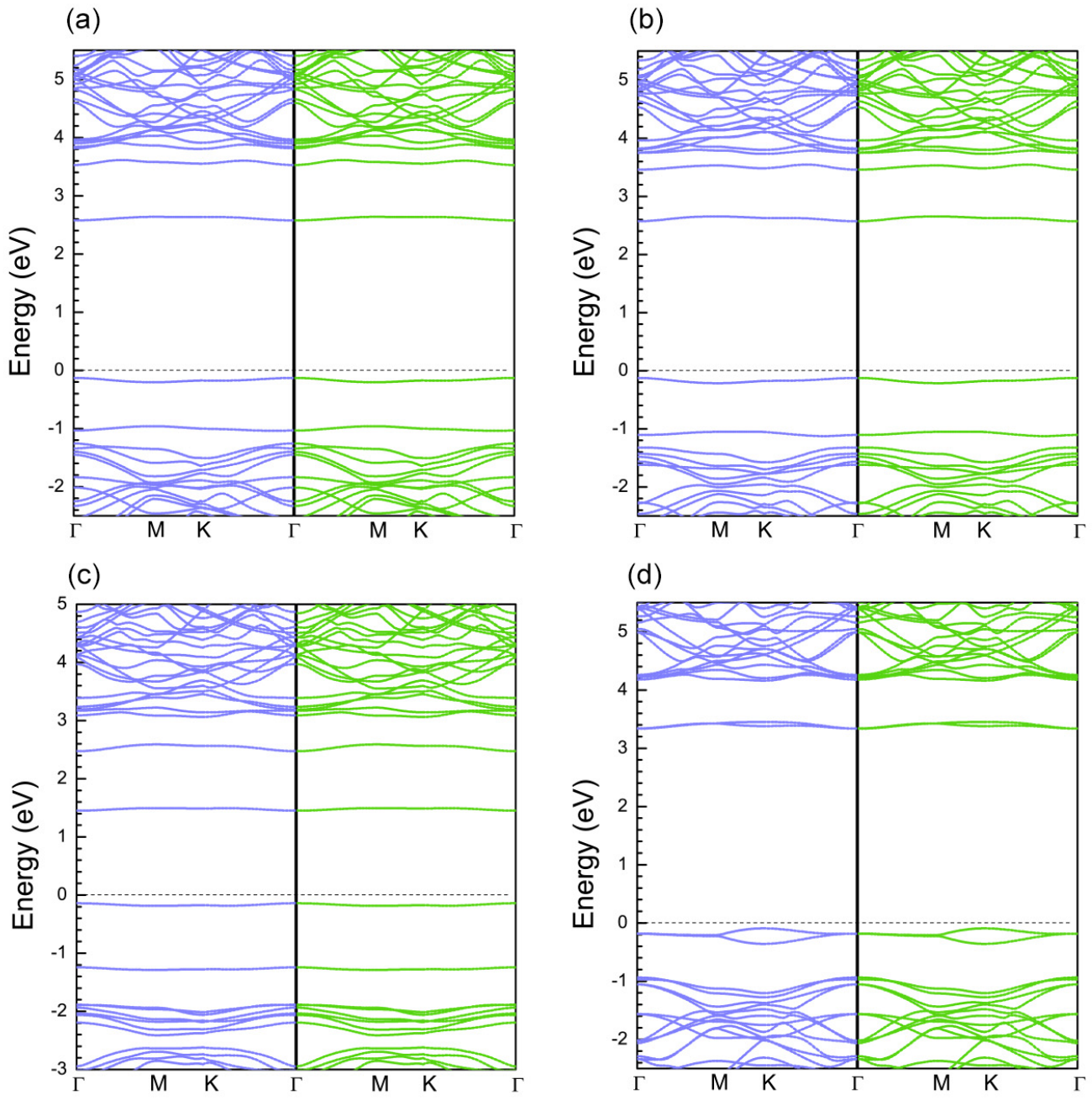

**Figure S1**. Spin-projected band structures of carbon clusters with equivalent B and N substitutions in hBN: **(a)** $C_{43}$, **(b)** $C_{44}$, **(c)** $C_{6H}$ and **(d)** $C_{6O}$.